\documentclass[a4paper,useAMS,usenatbib,usegraphicx]{mn2e}
\usepackage{times}
\usepackage{color}
\usepackage{amsmath}
\usepackage{amssymb}

\title[Sputtering in oblique C-type shocks]
{Sputtering in oblique C-type shocks}
\author[Van Loo et al.]
{S. Van Loo,$^1$\thanks{E-mail: svanloo@cfa.harvard.edu},
I.~Ashmore,$^2$, P. Caselli$^2$, S.~A.~E.~G.~Falle$^3$ and 
T.~W.~Hartquist$^2$ \\
$^1$ Harvard-Smithsonian Center for Astrophysics, 60 Garden St., 
Cambridge, Massachusetts 02138, USA \\
$^2$ School of Physics and Astronomy, University of Leeds, Leeds LS2 9JT, UK\\
$^3$ Department of Applied Mathematics, University of Leeds, Leeds LS2 9JT, UK}

\begin{document}

\date{Accepted -. Received -; in original form -}

\pagerange{\pageref{firstpage}--\pageref{lastpage}} \pubyear{}

\maketitle

\label{firstpage}

\begin{abstract}
We present the first results for the sputtering of grain mantles and cores 
obtained with self-consistent multifluid hydromagnetic models of C-type 
shocks propagating through dusty media. The threshold shock speed for mantle 
sputtering is about 10~km~s$^{-1}$ and is independent of density. The mantles 
are completely vapourised in shocks with speeds of 20-25~km~s$^{-1}$. At 
such shock speeds core sputtering commences and gas-phase SiO forms. Core 
destruction is not total in any C-type shock because grains are not 
completely destroyed in shocks with speeds near the minimum speeds at which 
J-type shocks appear. Due to the density-dependence of the critical shock speed
for this transition, higher gas-phase SiO fractional abundances are produced 
behind shocks propagating in lower density gas. For shock speeds near the 
threshold speeds for both core and mantle sputtering, sputtering is much 
greater for shock velocities at smaller angles relative to the upstream 
magnetic field. At higher shock speeds, the angular variation is still 
present but less pronounced.
\end{abstract}


\section{Introduction}\label{sec:intro}

A number of computational studies of sputtering in self-consistent models of
perpendicular C-type hydromagnetic shocks in dusty, dark regions have
been performed (e. g. \citealp{FPDF95}; \citealp{Sal97}; \citealp{Mal00}; 
\citealp{Gal08}; \citealp{Gu11}). However, the only numerical investigation 
of sputtering in oblique C-type shocks in such regions is not based on 
self-consistent shock models (\citealp{CHH97}).

\cite{PH94} attempted to construct a self-consistent, four fluid
model of an oblique C-type shock in which dust grains are described as a
charged fluid, as they are in the perpendicular C-type shock models of 
\cite{PHH90}. However, the integrations performed by \cite{PH94} failed to 
yield solutions corresponding to steady fast-mode shocks. \cite{W98} 
pointed out  that the failure is due to the saddle-point nature of the 
downstream state of a fast-mode dusty shock. \cite{F03} developed a 
numerical scheme for solving the time-dependent multifluid hydromagnetic 
equations that enables the construction of steady fast-mode solutions 
even when local equilibrium does not obtain. \cite{vLal09} have used 
the scheme along with non-equilibrium chemistry and treatments of 
grain charge and heating and radiative losses to investigate the evolution
of oblique C-type shocks propagating into media with constant upstream 
properties. \cite{Aal10} extended the applications to oblique C-type shocks
propagating through media with varying upstream properties.

Here we use models like those of \cite{vLal09} to study sputtering
in oblique C-type shocks in dusty, dark molecular regions. Section 2
contains descriptions of the assumptions, the physical processes included, 
the chemical  network, the evaluation of the sputtering rates and 
upstream boundary conditions. Results are presented in Section 3, and 
Section 4 contains a discussion of the results and conclusions.

\section[]{The Model}\label{sec:themodel}
The model includes neutral, ion, electron and two grain fluids. Flow 
variables depend on only the spatial coordinate $x$ and time $t$. The 
upstream magnetic field has components in the $x$ and $y$ directions, and 
the shock propagates in the $x$ direction.  Equations ~1 - 8 of 
\cite{vLal09} are solved with the scheme of \cite{F03}.
The inertia of the charged fluids, including the grain 
fluids, is neglected in this approach, so that their dynamics is solely
determined by the balance between the Lorentz force and collision forces 
with neutral particles. Additionally, the grain fluids are assumed to 
have zero pressure. The average charge on a grain is 
determined by the condition that the magnitude of the current of ions 
onto a grain equals the magnitude of the current of electrons onto a grain; 
Eqs.~10 and 11 of \citet[][see also 
Appendix~\ref{sect:appA}]{HHP87} are used to calculate these currents. 
The numerical method of \citet[][]{F03} provides a rigorous 
means to model steady shock structures and, as discussed by 
\citep[][]{vLal09} and \citet[][]{Aal10}, reproduces the results of, 
among others, \citet[][]{Del83}, \citet[][]{PH94}, \citet[][]{CW06} 
and \citet[][]{GPJ07}. Furthermore, although it neglects grain inertia, 
the method can accurately model neutral subshock structures \citep[][]{F03,Aal10}.

The evolution of the number density, $n(X)$, of chemical species $X$ 
is governed by a rate equation of the form:
\begin{eqnarray} \label{eqn:chemsource}
\frac{\partial n(X)}{\partial t} + \frac{\partial n(X) u_s}{\partial x} = 
s(X),
\end{eqnarray}
$u_{s}$ is the $x$ component of the ion or neutral velocity depending on 
whether $X$ is an ion or a neutral. $s(X)$ is the net rate of formation or 
destruction of species $X$ per unit volume per unit time due to the gas 
phase reactions and grain surface recombinations listed in 
Table~\ref{table:rateqns} and the return of material to the gas phase due 
to sputtering. Though no reactions involving SiO are shown in 
Table~\ref{table:rateqns}, atomic Si returned to the gas phase due to
sputtering is assumed to be converted to SiO immediately.

\begin{figure}  
  \includegraphics[width=\columnwidth]{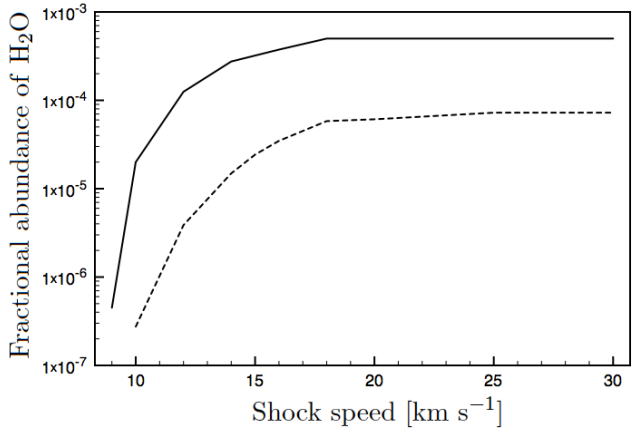}  
  \caption[water abundance]{The fractional abundance
        of H$_2$O produced by gas-phase chemistry (solid) and 
        mantle sputtering (dashed) as function of the shock speed
        for $n_{\rm H0} = 10^6 cm^{-3}$.}
  \label{fig:Fig4}
\end{figure}

\begin{table*}
\begin{minipage}{120mm}
  \caption{The chemical network. The column titled "Num." lists the 
     reaction number, that labelled "Reaction" gives the reaction partners 
     and products for each, that designated "Rate or Rate Coefficient" lists 
     the adopted rates and rate coefficients and that labelled "Ref." gives 
     the source of each coefficient: Source 1 is \protect \cite{PHH90}, 2 is 
     the UDFA database described by \protect \cite{Woel07}, 3 is 
     \protect \cite{Gal09} and 4 indicates that the value is calculated with 
     a routine in the code.}
  \label{table:rateqns}
  \begin{tabular}{|l|l|l|l|}
  \hline
  Num. & Reaction & Rate (s$^{-1}$) or Rate Coefficient (cm$^{3}$ s$^{-1}$) 
	& Ref. \\
  \hline
  1$^a$ & H$_{2}$ + CR $\rightarrow$ H$_{3}^{+}$ + e + CR & 3 $\times$ 
	10$^{-17}$& 1 \\ 
  2 & H$_{3}^{+}$ + CO $\rightarrow$ HCO$^{+}$ + H$_{2}$ & 1.7 $\times$ 
	10$^{-9}$ & 1, 2, 3 \\
  3 & H$_{3}^{+}$ + H$_{2}$O $\rightarrow$ H$_{3}$O$^{+}$ + H$_{2}$ & 
	5.9 $\times$ 10$^{-9}$ & 2, 3 \\
  4 & H$_{3}^{+}$ + grain$^{-}$ $\rightarrow$ H$_{2}$ + H + grain & 
	$\left(\Gamma_{ig} + \Gamma_{is}\right)\ n({\rm H}_{3}^{+})/n_{i}$ 
	& 4 \\
  5 & H$_{3}^{+}$ + O $\rightarrow$ H$_{3}$O$^{+}$ & 8.4 $\times$ 10$^{-10}$ 
	& 2, 3 \\
  6 & HCO$^{+}$ + e $\rightarrow$ H + CO & 2.76 $\times$ 10$^{-7} 
	\left({\rm T}_{e}/300\right)^{-0.64}$ & 3 \\
  7 & HCO$^{+}$ + Mg $\rightarrow$ H + CO + Mg$^{+}$ & 2.9 $\times$ 
	10$^{-9}$ & 2 \\
  8 & HCO$^{+}$ + H$_{2}$O $\rightarrow$ H$_{3}$O$^{+}$ + CO & 2.5 
	$\times$ 10$^{-9}$ & 2, 3 \\
  9 & HCO$^{+}$ + grain$^{-}$ $\rightarrow$ H + CO + grain & 
	$\left(\Gamma_{ig} + \Gamma_{is}\right)\ n({\rm HCO}^{+})/n_{i}$ & 4\\
  10 & Mg$^{+}$ + e $\rightarrow$ Mg + h$\nu$ & 2.8 $\times$ 10$^{-12} 
	\left({\rm T}_{e}/300\right)^{-0.86}$ & 1, 2 \\
  11 & Mg$^{+}$ + grain$^{-}$ $\rightarrow$ Mg + grain & 
	$\left(\Gamma_{ig} + \Gamma_{is}\right)\ n({\rm Mg}^{+})/n_{i}$ & 4\\
  12$^b$ & Mg$^{+}$ + H$_{2}$ $\rightarrow$ MgH$^{+}$ + H & 3.0 $\times$ 
	10$^{-10}$exp(-27854/T$_{\rm eff}$) & 1 \\
  13 & MgH$^{+}$ + e $\rightarrow$ Mg + H & 1.1 $\times$ 10$^{-7}$
	(T$_{e}$/300)$^{-0.5}$ & 1 \\
  14 & MgH$^{+}$ + grain$^{-}$ $\rightarrow$ Mg + H + grain & 
	$(\Gamma_{ig} + \Gamma_{is})\ n({\rm MgH}^{+})/n_{i}$ & 4 \\
  15$^c$ & H$_{3}$O$^{+}$ + e $\rightarrow$ H$_{2}$O + H & 3.29 
	$\times$ 10$^{-7}$(T$_{e}$/300)$^{-0.5}$  & 2, 3 \\
  16 & H$_{3}$O$^{+}$ + Mg $\rightarrow$ H$_{2}$O + Mg$^{+}$ + H & 
	1.8 $\times$ 10$^{-9}$ & 1 \\
  17 & H$_{3}$O$^{+}$ + grain$^{-}$ $\rightarrow$ H$_{2}$O + H + 
	grain & $(\Gamma_{ig} + \Gamma_{is})\ n({\rm H}_{3}{\rm O}^{+})/n_{i}$ 
	& 4 \\
  18$^d$ & O + H$_{2}$ $\rightarrow$ H$_{2}$O & 3.14 $\times$ 
	10$^{-13}$(T$_{n}$/300)$^{2.7}$exp(-3150/T$_{n}$) & 2, 3 \\
  \hline
  \end{tabular}
  \begin{flushleft}
  \noindent $^a$ Reaction 1 is an amalgamation of two reactions: 
	H$_{2}$ + CR $\rightarrow$ H$_{2}^{+}$ + e + CR 
	and H$_{2}^{+}$ + H$_{2}$ $\rightarrow$ H$_{3}^{+}$ + H where 
	CR denotes a cosmic ray. The adopted coefficient is 
	that of the first reaction since the proton transfer reaction 
	occurs sufficiently rapidly.
  
  \noindent $^b$ T$_{\rm eff}$ in the rate coefficient of Reaction 12 
	defines a weighted kinetic temperature at the ion-neutral 
	drift speed and is given by Eq.~43 of \cite{FPH85}.
  
  \noindent $^c$ Reaction 15 is an amalgamation of: \\
	H$_{3}$O$^{+}$ + e $\rightarrow$ OH + H$_{2}$ 
	$(6.02 \times 10^{-8} ({\rm T}_{e}/300)^{-0.5})$, \\
	H$_{3}$O$^{+}$ + e $\rightarrow$ OH + H + H 
	$(2.58 \times 10^{-7} ({\rm T}_{e}/300)^{-0.5})$ and \\
	H$_{3}$O$^{+}$ + e $\rightarrow$ H$_{2}$O + H 
	$(1.08 \times 10^{-8}({\rm T}_{e}/300)^{-0.5})$.  \\
	OH is not treated as a separate species, because the OH 
	produced by the first two reactions quickly reacts with 
	H$_{2}$ to yield H$_{2}$O (see $^d$).

  \noindent $^d$ Reaction 18 is also an amalgamation of two reactions: 
  	O + H$_{2}$ $\rightarrow$ OH + H and OH + H$_{2}$ $\rightarrow$ 
	H$_{2}$O + H. The coefficient adopted is that of the first 
	(slower) of the two.
  \end{flushleft}
\end{minipage}
\end{table*}

\begin{figure*}  
  \includegraphics[width=7.0in]{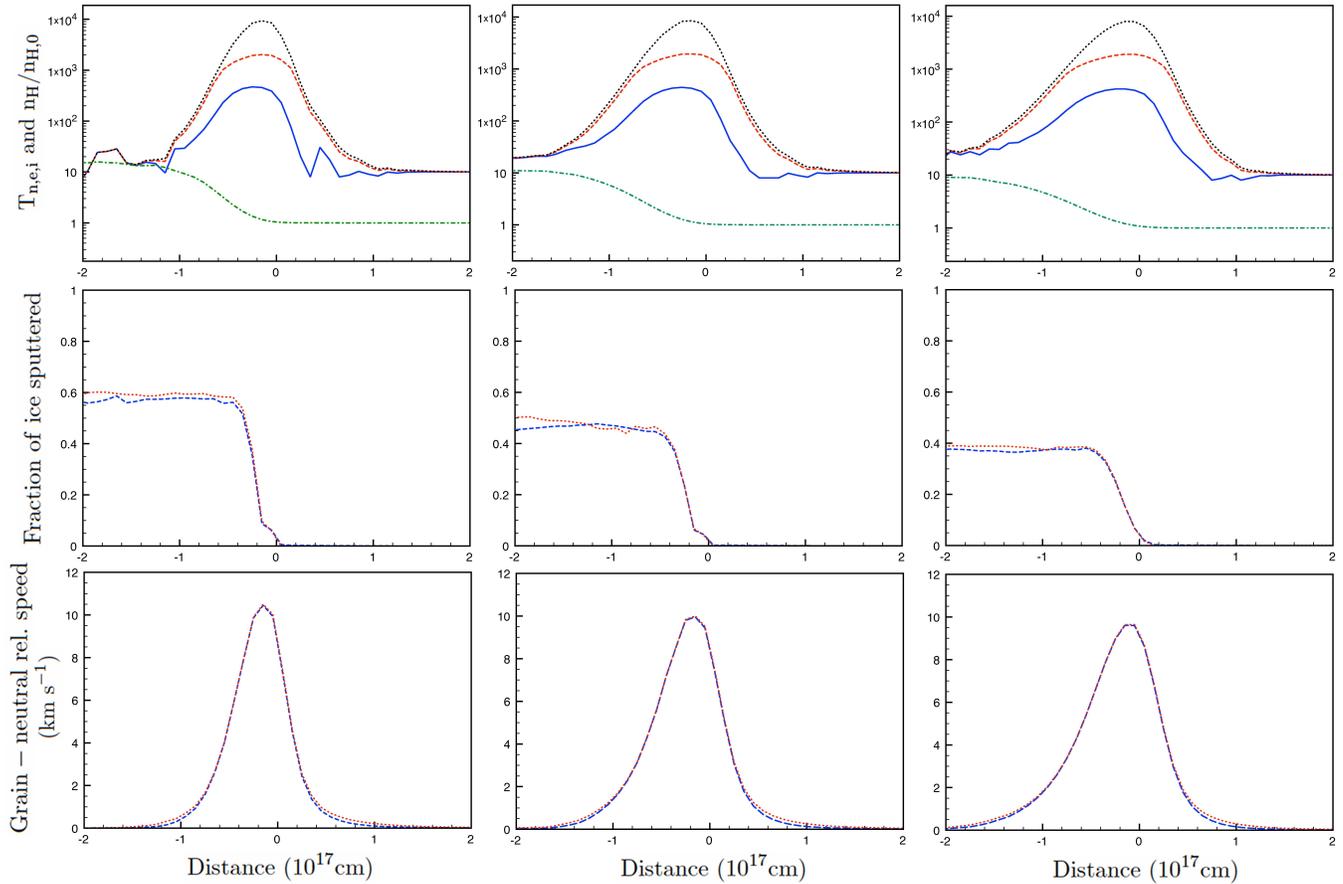}  
  \caption[Dependence of the sputtered mantle fraction on $\theta$ for 
	16 km s$^{-1}$ shocks and n$_{\rm H0}$ = 10$^{4}$ cm$^{-3}$.]{Fraction 
	of ice ejected from grain mantles for a shock speed of 16 km s$^{-1}$, 
	n$_{\rm H0}$ = 10$^{4}$ cm$^{-3}$ and $\theta$ of 30$^{\circ}$ (left), 
  	45$^{\circ}$ (middle) and 60$^{\circ}$ (right). The top row gives the 
	ion (black, dotted), electron (red, dashed) and neutral (blue, solid) 
	temperatures, as well as the neutral density normalised to its 
	upstream value (green, dash-dotted). The middle row shows the fraction 
	of ice ejected from large (blue, dashed) and small (red, dotted) 
	grains. The bottom row displays the speed of small (red, dotted) 
	and large (blue, dashed) grains relative to neutrals in km s$^{-1}$.} 
  \label{fig:obliquemantlesputter}
\end{figure*}

The model includes the radiative losses due to O, CO, H$_2$ and H$_2$O, which 
are calculated for the abundances of O and H$_2$O given by the integration of 
the chemical rate equations and the assumed constant fractional abundances of 
CO and H$_2$. Given those abundances, the cooling rates are evaluated as 
they were by \cite{vLal09}. They adopted the treatments of \cite{Del83} 
for the cooling of electrons and for OI emission, used data from \cite{Hel80} 
for the cooling of neutrals by H$_2$ and implemented the treatment of CO 
cooling due to \cite{HM79} and that of H$_2$O cooling developed by 
\cite{NM87}.

Sputtering due to H$_2$, CO, O, Mg, H$_2$O and SiO projectiles is included. 
As mentioned above, the return of material to the gas phase due to sputtering 
contributes to the source terms in the rate equations governing the gas 
phase chemistry; these contributions to the source terms are particularly 
important for the model gas phase abundances of Mg, H$_2$O and SiO. A grain 
consists of a refractory core made of olivine (MgFeSiO$_{4}$) and a mantle 
of water ice. A core remains intact until all of the ice has been sputtered. 
The sputtering rates are calculated as they were by \cite{JSal08}, who 
used Eq.~27 of \cite{DS79} for the sputtering rate. For all projectiles, they 
used Eq.~31 of \cite{DS79} to obtain the sputtering yields of ice at normal 
incidence. They used the same equation to obtain the corresponding yield 
for core material colliding with H$_2$.  However, for projectiles other 
than H$_2$, they employed the results of \cite{Mal00} for the sputtering 
yields of core material at normal incidence. Following \cite{DS79}, 
\cite{JSal08} took the angle averaged value of a yield to be twice the 
value of the normal incidence yield.

The upstream magnetic field strength is 
$B_0 = 1\mu {\rm G} (n_{\rm H0}/1~{\rm cm}^{-3})^{1/2}$, where n$_{\rm H0}$ 
is the upstream value of n$_{\rm H}$, the number density of hydrogen nuclei. 
A shock propagates with a speed v$_s$ at an angle $\theta$ with respect to 
the upstream magnetic field (for an oblique shock, 
$0^\circ \lneqq \theta \lneqq 90^\circ$). The grains have radii of 0.4 $\mu$m 
and 0.04 $\mu$m and masses of 8.03 $\times 10^{-13}$ g and 8.03 
$\times 10^{-16}$ g. In the upstream region, one percent of the mass is 
contained in grains with the number density of smaller grains 316 times 
that of the larger grains (i.e. the grains follow the grain-size distribution 
of \citet{MRN77}). 
As all the grains are silicate-based in our
model, the fractional abundance of Si, i.e. 5.8 $\times 10^{-5}$, is somewhat 
higher than the cosmic value of 3.37 $\times 10^{-5}$ \citep[][]{AG89}. 
The fractional abundance of CO to hydrogen nuclei has a 
constant value of 5 $\times$ 10$^{-5}$ ahead, throughout and behind the 
shock, and the upstream gas phase fractional abundances of O, Mg, H$_{2}$O 
and SiO are 4.25 $\times$ 10$^{-4}$, 10$^{-7}$, 0 and 0, respectively. 
Note that, including the fractional abundance of CO and of olivine, 
the total abundance of O in our model exceeds the cosmic abundance of O 
by less than a factor of two (the latest measured value being 
3.2 $\times 10^{-4}$ \citep{Meyeretal98} although \citet{Snow&Witt96} give 
4.6 $\times 10^{-4}$). The initial fractional ionisation is calculated 
as it was by \cite{vLal09}; HCO$^+$ and Mg$+$ are initially the only ions, 
but the chemistry leads to the production of significant fractional 
abundances of H$_{3}^{+}$ and H$_{3}$O$^{+}$ before heating by the shock 
occurs. Although upstream of a shock there is no water in the gas phase, 
the upstream fractional abundance of water ice molecules on the grains is 
7.25 $\times$ 10$^{-5}$ (cf. \citealp{WD91}); the upstream value of the 
ratio of the total mass of ice on small grains to the total mass of ice on 
big grains is taken to be 3.16.

We followed \cite{vLal09} by assuming that the initial flow corresponds to 
that of a J-type shock. The presented results are for the steady flows yielded 
by time integrations.

\begin{table*}
  \begin{minipage}{160mm}
  \caption{Fraction of mantle that is sputtered and gas phase water 
	fractional abundance enhancement due to sputtering. Results are 
	given for large and small grains and for both combined. 
        Although some small grain 
	sputtering does occur in the 9 km s$^{-1}$, n$_{\rm H0}$ = 
	10$^{6}$ cm$^{-3}$ case, the fraction of ice removed was 
	$< 10^{-4}$. }
  \label{table:mantlesputterfrac}
  \begin{tabular}{cccccccc}
  \hline
  & & \multicolumn{2}{c}{Large grains} & \multicolumn{2}{c}{Small grains} & 
	 \multicolumn{2}{c}{Large \& Small grains} \\
  \hline
  Shock Speed & n$_{\rm H0}$ & Fraction of & Contribution to &  Fraction of & 
	Contribution to &  Fraction of & Contribution to \\
  (km s$^{-1}$) & (cm$^{-3}$) & ice removed & downstream H$_{2}$O & ice 
	removed & downstream H$_{2}$O &  ice removed & downstream H$_2$O\\
  \hline
  10 & 1.00(4) & 1.20(-3) & 3.09(-8) & 3.80(-3) & 2.09(-7) &3.31(-3)&2.40(-7)\\
  12 & 1.00(4) & 4.00(-2) & 1.04(-6) & 5.50(-2) & 3.03(-6) &5.61(-2)&4.07(-6)  \\
  14 & 1.00(4) & 1.75(-1) & 3.05(-6) & 2.00(-1) & 1.10(-5) &1.94(-1)&1.41(-5) \\
  16 & 1.00(4) &  4.50(-1) & 7.83(-6) & 4.80(-1) & 2.64(-5) &4.72(-1)&3.42(-5)  \\
  20 & 1.00(4) & 1.00 & 1.74(-5) & 1.00 & 5.51(-5) &1.00&7.25(-5)\\
  \hline
  8 & 1.00(6) & 0.00 & - & 0.00 & - &0.0 &-\\
  9 & 1.00(6) & 0.00 & - & $<$1.00(-4) & $\approx$ 0.00 &$<$ 1.00(-4) & $\approx$ 0.00\\
  10 & 1.00(6) & 0.00 & - & 5.00(-3) & 2.75(-7)&3.79(-3) & 2.75(-7)\\
  12 & 1.00(6) & 2.50(-3) & 4.36(-8)  & 6.95(-2) & 3.83(-6)&5.34(-2) &3.87(-6) \\
  14 & 1.00(6) & 2.50(-2) & 4.36(-7) & 2.63(-1) & 1.45(-5)&2.06(-1) &1.49(-5) \\
  15 & 1.00(6) & 5.00(-2) & 8.71(-7) & 4.25(-1) & 2.34(-5) &3.35(-1) &2.43(-5)\\
  16 & 1.00(6) & 7.78(-2) & 1.36(-6) & 6.07(-1) & 3.34(-5) &4.79(-1) &3.48(-5)\\
  18 & 1.00(6) & 1.8(-1) & 3.14(-6) & 1.00 & 5.51(-5) &8.03(-1) &5.82(-5)\\
  20 & 1.00(6) & 3.4(-1) & 5.93(-6) & 1.00 & 5.51(-5) &8.42(-1) &6.10(-5)\\
  25 & 1.00(6) & 1.00 & 1.74(-5) & 1.00 & 5.51(-5) &1.00 &7.25(-5)\\
  \hline
  \end{tabular}
\end{minipage}
\end{table*}

\section{Results}\label{sec:results}
We focus on the abundances of H$_2$O and SiO as mantle and core sputtering 
releases these molecules into the gas phase. However, water is also produced 
by shock chemistry. To quantify the contributions of both processes 
we discuss the chemistry first and then grain sputtering. 

\subsection{Shock chemistry}
The specified initial relative abundances are not in steady state chemical 
equilibrium. The upstream gas phase ion abundances initially evolve on a 
timescale short compared to the flow time through the shock and then evolve 
further as the abundances of other key species (e. g. H$_2$O) vary. 
H$_{3}$O$^{+}$ and HCO$^{+}$ are the most common molecular ions, 
but the Mg$^{+}$ fractional abundance is about an order of magnitude greater 
than the total fractional abundance of molecular ions. The ratio of the total
atomic ion abundance to the total molecular abundance depends on the assumed
elemental abundance of metals. 

The assumed upstream water fractional abundance is about three orders of 
magnitude below the water fractional abundances obtained with standard 
ion-molecule chemical models, i.e. $\sim$ 10$^{-6}$ \citep[e.g.][]{Mal97,NLM95}.
However, the upstream value is close to the water abundances of $\sim$ 
10$^{-8}$ inferred for quiescent molecular gas towards Orion, M17 and other 
sources by \citet[][]{Bal00}. Recent Herschel observations of ortho-H$_{2}$O 
in a dark cloud \citep[][]{Casal10} imply values of $\sim$ 10$^{-10}$.
 
Within the shock front the gas heats up, and some reactions with 
barriers and endothermic reactions occur. Also, the heating of electrons leads
to enhanced electron impact rates on the grains, and the drift of the ions 
relative to the grains contributes further to efficient ion recombination
on grains, reducing the fractional ionisation within the shock. As usual, 
the primary sequence for H$_2$O formation in the shock involves the reactions 
of O and OH with H$_2$. Figure~\ref{fig:Fig4} shows the 
the fractional abundance of water produced in the shock as a function of 
the shock speed. At the lowest shock speed (i.e. 5 km s$^{-1}$), the 
shocked gas does not reach high enough temperatures to cause the conversion 
of much oxygen to water. At a shock speed of 10~km~s$^{-1}$ a few percent of 
the gas phase oxygen is converted to water, but it is only at shock speeds 
close to 20~km~s$^{-1}$ and above at which the atomic oxygen abundance falls 
by several orders of magnitude, and nearly all of the gas phase oxygen is 
in water. The H$_2$O abundance is then a few times 10$^{-4}$. 

\subsection{Mantle sputtering}
Within a C-type shock, grains experience high-speed impacts with gas 
particles. Sputtering leads to a continuous erosion of the icy grain mantle 
by which water is returned to the gas phase. 

Grain mantle sputtering occurs for impact speeds $> 10/\sqrt{\mu}$~km~s$^{-1}$ 
with $\mu$ the mean mass of the impacting species in atomic mass units 
\citep[e.g.][]{Mal00}. Since CO is the most massive abundant projectile with 
$\mu = 28$, mantle sputtering starts for grain-neutral drift speeds as 
low as $\approx 2$~km~s$^{-1}$. (Note from Fig.~\ref{fig:obliquemantlesputter}
that appreciable sputtering only occurs at drift speeds above 6 km~s$^{-1}$). 
This value of the drift speed is independent of the grain size, because the 
mantle molecules are bound with the same energy. However, the threshold drift 
speed is reached at different shock speeds for the small and large grains. 
Small grains have Hall parameters\footnote{The Hall parameter is the ratio 
between the gyrofrequency of a charged particle and the collisional momentum 
transfer frequency of the charged particle with neutrals.} that are large 
enough that they move with the ions and electrons. The Hall parameter for 
the large grains is above unity but sufficiently small that they are only 
moderately coupled to the magnetic field and, thus, move at velocities between 
those of the electrons and the neutrals. Consequently, the grain-neutral drift
speed for the large grains is smaller than for the small grains at a given 
shock speed.

\begin{figure}  
  \includegraphics[width=\columnwidth]{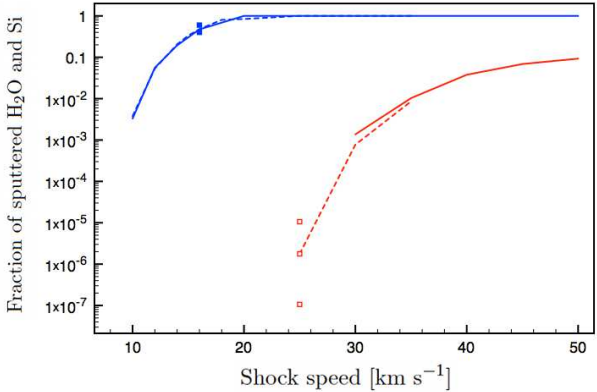}  
  \caption[Sputtering fractions]{The fraction of mantle ice 
        (blue) and core silicon (red) released in the gas-phase by 
        sputtering as function of shock speed for values 
	of $n_{\rm H0}$ of 10$^4$ cm$^{-3}$ (solid) and 10$^6$ cm$^{-3}$ 
        (dashed). The squares show the sputtering fractions for 
        shock angles of 30, 45 and 60 degrees for selected shock speeds;
	the sputteringefficiency decreases with increasing angle.}
  \label{fig:Fig3}
\end{figure}

No  projectile induces noticeable sputtering when v$_{s}$ = 5 km s$^{-1}$, but 
when v$_{s}$ has a higher value, small grains are sputtered.  Table 
\ref{table:mantlesputterfrac}, as well as Fig.~\ref{fig:Fig3},
displays the fraction of the initial mantle ice 
that is sputtered by a shock for which $\theta = 45^{\circ}$. Results are 
given for various values of the shock speed and values of n$_{\rm H0}$ of 
10$^{4}$ cm$^{-3}$ and 10$^{6}$ cm$^{-3}$. It also gives the enhancement in 
the downstream gas phase water fractional abundance due to sputtering. If we 
define the sputtering threshold shock speed as the shock speed for which 1\% 
of the ice is removed \citep[cf.][]{CHH97}, the threshold shock speed for 
small grain sputtering is 10~km~s$^{-1}$ and for large grain sputtering it is 
10 (12) km~s$^{-1}$ for an upstream number density of 10$^4$ 
(10$^6$)~cm$^{-3}$. 

One also sees that the fraction of material sputtered from small grains is 
not very sensitive to n$_{\rm H0}$, but particularly at lower shock speeds 
the fraction of material sputtered from the large grains does depend on 
n$_{\rm H0}$. This again is a result of the coupling of the grains to the 
magnetic field (see above). The large grains have a larger Hall parameter 
in the lower density regime and, thus, have a larger neutral-grain drift 
speed than in the high-density regime. 

For shocks for which $\theta = 45^{\circ}$, the highest sputtering
rate per unit time per unit volume of mantles is due to collisions with CO 
for shock speeds less than about 10 km s$^{-1}$ and to collisions with H$_2$O 
for speeds above about 14 km s$^{-1}$. At intermediate speeds, collisions 
with O also contribute non-negligibly as the relative importance of CO and 
H$_2$O collisions varies. Above 30~km~s$^{-1}$, H$_2$ dominates the sputtering 
of both grain types. However, at these shock speeds, the icy mantles of both 
grain species are completely eroded. The contribution of the grain 
sputtering to the water abundance is even then negligible compared to the 
contribution of the gas-phase oxygen conversion (see 
Fig.~\ref{fig:Fig4}). 

\begin{table*}
\begin{minipage}{152mm}
  \caption{Fraction of core elemental silicon that is sputtered and gas phase
  	SiO fractional abundance enhancement due to sputtering.
  	Results are given for large and small grains and for 
        both combined. }
  \label{table:coresputterfrac}
  \begin{tabular}{cccccccc}
  \hline
  & & \multicolumn{2}{c}{Large grains} & \multicolumn{2}{c}{Small grains} & 
			\multicolumn{2}{c}{Small \& large grains} \\
  \hline
  Shock Speed & n$_{\rm H0}$ & Fraction of &  Contribution to & Fraction of 
	& Contribution to & Fraction of & Contribution to\\
  (km s$^{-1}$) & (cm$^{-3}$) & Si removed &  downstream SiO & Si removed 
	& downstream SiO & Si removed & downstream SiO\\
  \hline
  20 & 1.00(4) & 0.00 & - & $<$1.00(-4) & $\approx$0.00 &$<$ 1.00(-4) &$\approx$ 0.00\\
  30 & 1.00(4) & 3.62(-4) & 1.60(-8) & 4.58(-3) & 6.40(-8) &1.37(-3) &8.00(-8)\\
  35 & 1.00(4) & 2.83(-3) & 1.25(-7) & 3.40(-2) & 4.75(-7) &1.03(-2) & 6.00(-7)\\
  40 & 1.00(4) & 5.59(-3) & 2.47(-7) & 1.40(-1) & 1.95(-6) &3.78(-2) &2.20(-6)\\
  45 & 1.00(4) & 1.12(-2) & 4.94(-7) & 2.51(-1) & 3.51(-6) &6.87(-2) &4.00(-6)\\
  50 & 1.00(4) & 2.60(-2) & 1.15(-6) & 3.04(-1) & 4.25(-6) &9.28(-2) &5.40(-6)\\
  \hline
  20 & 1.00(6) & 0.00 & - & 0.00 & - & 0.00 & -\\
  25 & 1.00(6) & 0.00 & - & 7.37(-6) & 1.03(-10) & 1.77(-6) & 1.03(-10)\\
  30 & 1.00(6) & 0.00 & - & 3.22(-3) & 4.50(-8) & 7.73(-4) & 4.50(-8)\\
  35 & 1.00(6) & 9.76(-4) & 4.31(-8) & 3.26(-2) & 4.56(-7) & 8.57(-3) & 4.99(-7)\\
  \hline
  \end{tabular}
\end{minipage}
\end{table*}

\subsection{Core sputtering}
Once the grain mantles are vapourised, the grain cores begin to erode as 
long as the impact speeds of the projectiles are high enough. As the binding 
energy of a particle in the grain core is higher than in the mantle, higher 
grain-neutral drift speeds are necessary. Again CO is the sputtering 
projectile of most importance at lower speeds, and small grains start to 
contribute to the SiO gas-phase abundance at shock speeds of 25~km~s$^{-1}$ 
(see Fig.~\ref{fig:Fig3} for shocks for which 
$\theta = 45^{\circ}$). This result is similar to 
the one of \citet[][]{CHH97} and \citet[][]{Mal00}. Collisions with CO 
dominate the sputtering at shock speeds up to 30 km s$^{-1}$. Above shock 
speeds of about 35 km s$^{-1}$ the highest sputtering rate per unit time 
per unit volume of elemental silicon is due to collisions with H$_2$O.  
Table \ref{table:coresputterfrac} displays the fraction of the initial core 
elemental silicon that is sputtered. It also gives the enhancement in the 
downstream gas phase SiO fractional abundance due to sputtering.

Similarly to the results for mantle sputtering, the core sputtering of the 
small grains is independent of $n_{\rm H0}$ (as their Hall parameter is much 
larger than unity). The large grains are more efficiently sputtered in the 
low-density regime but contribute only a negligible amount to the SiO 
gas-phase abundance. For a range of shock speeds, the SiO production rises 
with increasing speed with values similar as in Fig.~2 of 
\citet[][]{Gal08a}. However, the SiO abundance reaches an upper limit as 
C-type shocks do not exist for shock speeds above a certain value. For 
higher shock speeds, shocks are of J-type. For $n_{\rm H0}=10^4~{\rm cm^{-3}}$ 
the transition occurs for a shock speed between 50 and 55~km~s$^{-1}$, while 
a J-type shock occurs for a minimum shock speed between 35 and 40~km~s$^{-1}$
for $n_{\rm H0} = 10^6~{\rm cm^{-3}}$. At these speeds the grain cores are not 
completely destroyed, i.e. only a few percent of the core is removed in the 
high-density case and about a third of the core is destroyed in the 
low-density regime. Hence, the maximum SiO fractional abundance differs by 
an order of magnitude between the two density regimes.  

\subsection{Angular dependence}
Up to now we have discussed the grain sputtering for an oblique shock for 
which $\theta = 45^\circ$. However, \citet[][]{vLal09} showed that the shock 
properties, such as the shock width, change as function of $\theta$. Therefore,
variations, with angle, of the H$_2$O and SiO abundances due to sputtering is 
expected. 

\begin{figure*}  
  \includegraphics[width=7.0in]{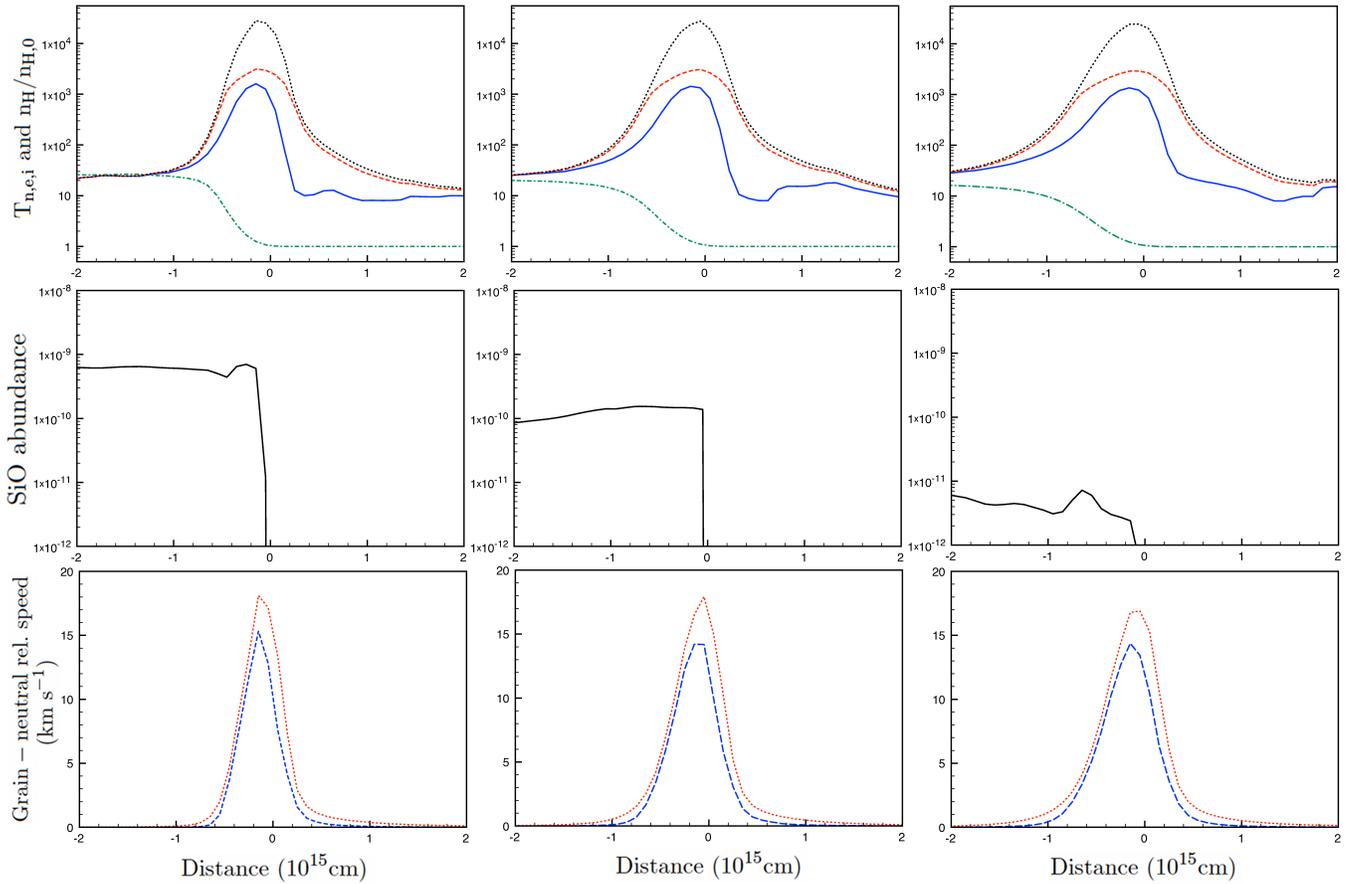}
  \caption[Dependence of the gas-phase SiO abundance on $\theta$ for 25 km 
	s$^{-1}$ shocks and n$_{\rm H0}$ = 10$^{6}$ cm$^{-3}$.]{Gas phase 
	fractional abundance of SiO for a shock speed of 25 km s$^{-1}$, 
	n$_{\rm H0}$ = 10$^{6}$ cm$^{-3}$ and $\theta$ of 30$^{\circ}$ (left 
	column), 45$^{\circ}$ (middle column) and 60$^{\circ}$ (right column). 
  	The top row gives the ion (black, dotted), electron (red, dashed) and 
	neutral (blue, solid) temperatures, as well as the neutral density 
	normalised to its upstream value (green, dash-dotted). The middle row 
	shows the gas phase fractional abundances of SiO. The bottom row 
	displays the speed of small (red, dotted) and large (blue, dashed) 
	grains relative to neutrals in km s$^{-1}$.}  
  \label{fig:obliquecoresputter}
\end{figure*}

Figure \ref{fig:obliquemantlesputter} presents results for 16 km s$^{-1}$ 
shocks propagating into media with n$_{\rm H0}$ = 10$^{4}$ cm$^{-3}$ at angles 
of 30$^{\circ}$, 45$^{\circ}$ and 60$^{\circ}$ with respect to the upstream 
magnetic field. The fraction of the ice that is sputtered from each type of 
grain, rather than the gas phase abundance of water, is shown because much of 
the gas phase water exists due to the conversion of upstream gas phase atomic 
oxygen to water. The results in Fig.~\ref{fig:obliquemantlesputter} indicate 
that the H$_2$O abundance is lowest in the shock with the largest width. It 
is obvious that the shock width is irrelevant for the level of sputtering as 
the sputtering depends mainly on the neutral-grain drift speeds. As the bottom 
row of Fig.~\ref{fig:obliquemantlesputter} shows, the peak small and large
grain-neutral drift speeds are smallest at large angles; they are 
10.5~km~s$^{-1}$ at 30$^{\circ}$, 10~km~s$^{-1}$ at 45$^{\circ}$ and 
9.5~km~s$^{-1}$ at 60$^{\circ}$. This behaviour is consistent with the results 
shown in Fig.~2 of \cite{vLal09} who found that the $y$ and $z$ components of 
the velocities of charged species are greater for smaller angles.  Furthermore,
the length scale over which the drift speeds are higher than 6~km~s$^{-1}$,
i.e. the minimum drift speed for which mantle sputtering is efficient, are 
similar for all angles. The angular variations in the fraction of water that 
is sputtered are significant and range from 0.4 at 60$^\circ$ to 0.6 at 
30$^\circ$  Such variations disappear with increasing shock speed because for 
high enough speeds vaporisation of the grain mantle is complete for all angles.

Figure \ref{fig:obliquecoresputter} presents results for 25 km s$^{-1}$ shocks 
propagating into media with n$_{\rm H0}$ = 10$^{6}$ cm$^{-3}$ at angles of
30$^{\circ}$, 45$^{\circ}$ and 60$^{\circ}$ with respect to the upstream medium.
The results in Fig.~\ref{fig:obliquecoresputter} show that the SiO abundance 
too is lowest in the shock with the largest width. Again these differences can 
be attributed to the peak small grain-neutral drift speeds which are smallest 
at large angles; they are 18.2 km s$^{-1}$ at 30$^{\circ}$, 18 km s$^{-1}$ at 
45$^{\circ}$ and 17 km s$^{-1}$ at 60$^{\circ}$. Contrary to the H$_2$O 
results, these modest differences in drift speeds have large effects on the 
amount of elemental silicon that is sputtered, i.e. there is a difference of 
two orders of magnitude between the SiO abundance at 30$^\circ$ and 60$^\circ$.
However, note that the shock speed is close to the threshold speed for core 
sputtering. Then small variations in the drift speed can mean the difference 
between core sputtering or no sputtering at all. For higher shock speeds the 
angular differences in SiO abundance are less pronounced, but the trend of 
lower abundances for larger $\theta$ remains. 

\section{Discussion and Conclusions}
In this paper we present the first self-consistent multifluid MHD simulations 
of grain mantle and core sputtering in oblique C-type shocks propagating 
through dusty media. The small grains are more tightly coupled to the magnetic 
field than the large grains with a strong coupling in both low- and 
high-density regimes. This maximises the drift speeds between the small grains 
and the neutral sputtering projectiles resulting in more efficient sputtering 
of the small grains. Mantle sputtering starts at shock speeds of about 
10~km~s$^{-1}$, and the mantles are completely vapourised for shock speeds 
around 20-25~km~s$^{-1}$. The sputtering of the icy mantles returns water back 
to the gas phase. However, neutral-neutral chemistry dominates the production 
of gas phase water by an order of magnitude. 

Once the mantles are completely eroded, core sputtering commences and releases 
silicon to form SiO. The abundance of SiO in the gas phase never saturates as 
even at the critical speed for the transition from C- to J-type shocks not all 
of a grain core is destroyed. While the critical speed is 50-55~km~s$^{-1}$ 
for $n_{\rm H0} = 10^4~{\rm cm^{-3}}$, it is only 40~km~s$^{-1}$ for 
$n_{\rm H0} = 10^6~{\rm cm^{-3}}$. For a given shock speed the efficiency of 
small grain sputtering is not sensitive to the density. Consequently, the 
density dependence of the C- to J-type transition means that higher SiO 
abundances can be produced by shocks in the low-density regime. 

An important result is that the sputtering yields depend on the angle 
between the magnetic field and the shock normal. Especially, at the threshold 
drift speeds for mantle or core sputtering, small angular variations can induce
large changes in abundances. For example, for shocks propagating at 
25~km~s$^{-1}$ in media with number densities of 10$^6 {\rm cm^{-3}}$, the 
SiO abundance differs by two orders of magnitude for $\theta = 30^\circ$ and 
$60^\circ$. This has important consequences for the interpretation of 
observations. For example, discrete emission features in SiO observations have 
been ascribed  to different shocks being present within the telescope beam 
\citep[e.g.][]{Izas2009}. However, such features can be explained by a bow 
shock interacting with an inhomogeneity in the medium. The flanks of the bow 
shock have lower speeds than the head and also make different angles with 
the magnetic field. In an inhomogeneous medium some regions of the bow-shock 
are then highlighted resulting in discrete SiO emission features. Such an 
assumption, however, needs to be further examined by extending the planar 
shock model to multiple dimensions. 

Direct comparisons of the results with those of other researchers is 
complicated as the only previous work (\citealp{CHH97}) on sputtering in 
oblique shocks is not based on self-consistent shock models. Also, the 
treatment of core sputtering in that work differed from that developed by 
\cite{Mal00} and used in subsequent work. However, the results given in 
Table \ref{table:coresputterfrac} for the fraction of elemental silicon 
sputtered from small grains are in harmony with the results of \citet{CHH97} 
for a shock for which $\theta = 45^\circ$. Other self-consistent studies 
concern only perpendicular shocks ($\theta = 90^\circ$) and, thus, should 
find similar, although lower, values for shock speed exceeding
the threshold value for sputtering. \cite{Gal08a}, who assumed a somewhat 
larger upstream magnetic field strength and different grain population, 
found that the fraction of silicon that is sputtered is about 0.04 for 
n$_{\rm H0} = 10^{4}$ cm s$^{-1}$ and a shock speed 40 km s$^{-1}$. 
\cite{Gu11} who adopt a kinetic method for the dust grain dynamics in favour
of a fluid approach, find a value of 0.02 for the same model as \citet[][]{Gal08a}.  
The corresponding result given in Table \ref{table:coresputterfrac} is 0.04.

The SiO abundances found in our simulations only represent upper limits. 
Observations indicate that silicate-based grains only account for about 
75\% of the interstellar dust mass with the rest of the mass stored in 
carbonaceous grains or PAHs \citep[][]{Draine2011}. Furthermore, the 
conversion of atomic silicon into SiO is not fully efficient above densities 
of 10$^5 {\rm cm^{-3}}$ which possibly reduces the SiO abundance due to 
sputtering by an order of magnitude \citep[][]{Gal08}.

Unlike \cite{Gal08a} and \cite{Gu11}, who included grains 
40 times smaller than the smallest 
that we have assumed to be present, we have not considered the destruction of 
grain material due to grain-grain collisions. Our choice to focus solely on 
sputtering first was based in part on a desire to understand the effects 
of obliqueness on sputtering before looking at the complicated way in which 
obliqueness, sputtering and grain-grain collisions all act together to destroy 
grains. Our decision to do so was informed by the estimates of \cite{CHH97} 
that showed for grain populations of the sizes that we selected, grain-grain 
collisions would contribute significantly to mantle removal only at shock 
speeds near the threshold even for high pre-shock densities. Our choice has 
allowed us to see that obliqueness affects sputtering near the threshold shock 
speed significantly. However, considering the findings of both \cite{CHH97} 
and of \cite{Gu11} for the return of silicon to the gas phase, as well as the 
results given in Figure \ref{fig:obliquecoresputter}, we conclude that the 
future inclusion of both grain-grain collisions and obliqueness in shock 
models is necessary to infer the threshold shock speed for the injection of 
significant quantities of silicon into the gas phase.

\section*{Acknowledgments}
We thank the anonymous referee for his/her constructive 
comments.
The authors are grateful for the support provided by a studentship and 
successive rolling grants awarded by the UK Science and Technology Funding 
Council and from the SMA Postdoctoral Fellowship of the Smithsonian 
Astrophysical Observatory held by SVL.

\begin{appendix}
\section{Particle current onto a grain}\label{sect:appA}
\citet{HHP87} derived general expressions for the current of charged 
particles (with charge $Z_j$) onto a dust grain carrying a charge of $Z_g$. 

For $Z_g Z_j \leq 0$, 
\begin{align*}
   \Gamma_{jg} &= \pi a^2 n_j Z_j e \frac{c_j^2}{2v_{jg}} 
     \Bigl\{\left(1+2\left(\frac{v_{jg}}{c_j}\right)^2 - 2Z_jZ_g\xi \right) \\
    &\times {\rm erf} \left(\frac{v_{jg}}{c_j}\right) + \frac{2(v_{jg}/c_j)}{\sqrt{\pi}} 
	\exp\left(-\left(\frac{v_{jg}}{c_j}\right)^2 \right) \Bigr\},
\end{align*}
and, for $Z_g Z_j \ge 0$, 
\begin{align*}
    \Gamma_{jg} &= \pi a^2 n_j Z_j e \frac{c_j^2}{2v_{jg}} 
     \Bigl\{\frac{1}{2}\left(1+2\left(\frac{v_{jg}}{c_j}\right)^2 - 2Z_jZ_g\xi \right) \\
    &\times \left[{\rm erf}\left(\frac{v_{jg}}{c_j} + \sqrt{Z_jZ_g\xi}\right) + 
     {\rm erf}\left(\frac{v_{jg}}{c_j} - \sqrt{Z_jZ_g\xi}\right) \right] \\
    & + \frac{1}{\sqrt{\pi}} \left(\frac{v_{jg}}{c_j}+ \sqrt{ZjZ_g\xi} \right) 
	\exp\left(-\left(\frac{v_{jg}}{c_j}-\sqrt{ZjZ_g\xi}\right)^2\right) \\
    & + \frac{1}{\sqrt{\pi}} \left(\frac{v_{jg}}{c_j}- \sqrt{ZjZ_g\xi} \right) 
	\exp\left(-\left(\frac{v_{jg}}{c_j}+\sqrt{ZjZ_g\xi}\right)^2\right) \Bigr\},
\end{align*}
where $a$ is the grain radius, $n_j$ the number density of the charged  
fluid, $v_{jg}$ the relative drift velocity between the grain and 
the charged fluid, $c_j$ the sound speed of the charged fluid and 
$\xi = e^2/ak_BT$. The charged fluid can be the ion or electron fluid.
\end{appendix}

\label{lastpage}

\end{document}